\documentstyle[12pt]{article}

\def\ffrac#1#2{\textstyle{#1\over#2}\displaystyle}
\def\zb{{\bar z}}
\def\Tt{{\widetilde{\cal T}}}
\def\qb{{\bar q}}
\def\Phit{{\widetilde\Phi}}

\begin{document}
\baselineskip=18pt
\pagestyle{empty}
\vspace{-1mm}
\begin{flushright}
cond-mat/0111031
\end{flushright}
\vspace{5mm}
\begin{center}
{\bf \LARGE The Stress Tensor in Quenched Random Systems\\}
\vspace{8mm}
{\bf \large John Cardy\\}
\vspace{2mm}
{Department of Physics\\
Theoretical Physics\\1 Keble Road\\Oxford OX1 3NP, UK\\
\& All Souls College, Oxford\\}
\end{center}
\vspace{6mm}
\begin{abstract}
The talk describes recent progress in understanding the behaviour of
the stress tensor and its correlation functions at a critical point of
a generic quenched random system.
The topics covered include:
(i) the stress tensor in random systems considered
as deformed pure systems;
(ii) correlators of the stress tensor at a random fixed point:
expectations from the replica approach and $c$-theorem sum rules;
(iii) partition function on a torus;
(iv) how the stress tensor enters into correlation functions:
subtleties
with Kac operators.
\end{abstract}
\vfill
\noindent$^*$ Talk presented at Workshop on Statistical Field Theory,
Como, Italy, June 2001.
\newpage

\pagestyle{plain}
\setcounter{page}{1}
\setcounter{equation}{0}

This talk is about the stress tensor in generic quenched random systems
in which we expect the quenched averaged correlation functions to be
those of some conformal field theory.
Many of the results generalise to arbitrary dimension, but I shall
take $d=2$ for simplicity.

\subsection*{Quenched random systems as deformed pure systems.}
Consider the quenched random system defined by the action
\begin{displaymath}
S=S_P+\sum_{i=1}^N\int h_i(r)\Phi_i(r)d^2\!r
\end{displaymath}
$S_P$ is a non-random CFT; $\Phi_i(r)$ is a set of primary fields
(assumed to be scalars -- we can generalise to vector couplings).
The $h_i(r)$ are quenched random variables, with 
$\overline{h_i(r)}=0$ and
$\overline{h_i(r')h_j(r'')}=\lambda_{ij}\delta(r'-r'')$.
We take $N=1$ for simplicity in this talk.

We are interested in the RG flow: $S_P\Longrightarrow$(random fixed point).
The perturbation is not necessarily small: the idea is to see
how objects in $S_P$ \em deform \em in the full theory. One of the tools
will be replica group theory. There is an analogy with the use of group
theory in atomic physics, where we can deduce the nature of the the
splittings in the spectrum even when the couplings are relatively
large.

Recall how Zamolodchikov considered
deformed CFT in pure systems: the action is 
\begin{displaymath}
S=S_0-\lambda\int\Phi(r)d^2\!r
\end{displaymath}
where $\lambda$ is a constant. The deformation of the $zz$ component of
the stress tensor is, to first order in $\lambda$,
\begin{displaymath}
\delta T(z,\bar z)=
\lambda\int_{|z'-z|>a}d^2\!z'T(z)\cdot\Phi(z',\zb')
\end{displaymath}
so that the conservation equation becomes
\begin{eqnarray*}
\partial_\zb T(z,\bar z)&=&
\lambda\int d^2\!z'\delta(|z-z'|^2-a^2)(z-z')\\
&&\qquad\left({\Delta\over(z-z')^2}\Phi(z,\zb)+{1-\Delta\over(z-z')}\partial_z
\Phi(z,\zb)+\cdots\right)\\
&=&-\partial_z\Theta
\end{eqnarray*}
where 
\begin{eqnarray*}
\Theta&=&-\pi\lambda(1-\Delta)\Phi\qquad  (d=2)\\
&&\propto-\lambda(d-x_\Phi)\Phi\qquad  ({\rm general\ } d)
\end{eqnarray*}

Note that no higher order terms in $\lambda$ arise, as long as 
no additional renormalisation is required. This is unnecessary if
$x_\Phi<d$), but in general
$\Theta\propto \beta(\lambda)\Phi_R$, so that $\Theta$ vanishes at the
new IR fixed point.

Now do this for a \em random \em coupling $\lambda\to h(z,\zb)$:
\begin{eqnarray*}
\partial_\zb T&=&
\int d^2\!z'\delta(|z-z'|^2-a^2)(z-z')h(z',\zb')\\
&&\qquad\left({\Delta\over(z-z')^2}\Phi(z,\zb)+{1-\Delta\over(z-z')}\partial_z
\Phi(z,\zb)+\cdots\right)
\end{eqnarray*}
but now $h(z',\zb')$ is white noise. After some stochastic calculus,
the result is
\begin{displaymath}
\partial_\zb T+\partial_z\Theta=K
\end{displaymath}
where
\begin{eqnarray*}
\Theta(z,\zb)&=&-\pi(\ffrac12-\Delta_\Phi)h(z,\zb)\Phi(z,\zb)\qquad(d=2)\\
&&[\propto(d-\ffrac d2-x_\Phi)h\Phi]\qquad  ({\rm general\ } d)\\
{\rm and\ }\qquad K&=&\ffrac12\pi(h\partial_z\Phi-\Phi\partial_z h)
\end{eqnarray*}

Note that $T$ and $\Theta$ are the components of the stress tensor for 
a \em given realisation \em of randomness, not the quenched average!
The extra contribution
$\ffrac d2$ comes from the white noise nature of $h(r)$.
The derivative
$\partial h$ does not make literal sense since $h(r)$ is a stochastic
function: it is interpreted by integrating by parts in correlators.
There is a similar
equation relating $\overline T$ to the \em same \em $\Theta$:
so that $T_{z\zb}=T_{\zb z}\propto\Theta$: the stress tensor is
symmetric even in the random system, because local
rotational symmetry is preserved (The results are slightly modified if the
coupling is to a random vector).

\subsection*{Replica formulation.}

How does all this appear within the replica formulation?
\begin{eqnarray*}
\overline{{Z^n}}&=&\int{\cal D}h e^{-(1/2\lambda)\int h^2d^2\!r}
{\rm Tr}\,e^{-\sum_a S_{P,a}+\int h\sum_a\Phi_a d^2\!r}\\
&=&{\rm Tr}\,\int{\cal D}h
e^{-(1/2\lambda)\int(h-\lambda\sum_a\Phi_a)^2d^2\!r}
e^{-\sum_a S_{P,a}+\frac12\lambda\int\sum_{a\not=b}\Phi_a\Phi_bd^2\!r}
\end{eqnarray*}
which has the form of a translationally invariant perturbed CFT.

The replicated theory has a stress tensor $\cal T$ which is a deformation
of $\sum_a T_a$, so
\begin{displaymath}
\partial_\zb{\cal T}+\partial_z\vartheta=0
\end{displaymath}
where $\vartheta=-\frac12\lambda\pi(1-2\Delta)\sum_{a\not=b}\Phi_a\Phi_b$.
Note that neither $\cal T$ nor $\vartheta$ are the components of the 
true stress tensor, discussed in the previous section.

At the new fixed point, $\vartheta=0$, and 
\begin{displaymath}
\langle{\cal T}(z){\cal T}(0)\rangle=c(n)/2z^4
\end{displaymath}
where, by the $c$-theorem sum rule,
\begin{displaymath}
c(n)-nc_P=-(12/\pi)\int
r^2\langle\vartheta(r)\vartheta(0)\rangle_cd^2\!r
\end{displaymath}

This has the following interpretation:
\begin{eqnarray*}
\langle{\cal T}{\cal T}\rangle&=&\sum_{a,b}
\langle T_aT_b\rangle=n\langle T_1T_1\rangle+n(n-1)\langle
T_1T_2\rangle\\
&\sim&n\left(\overline{\langle TT\rangle}-\overline{\langle T\rangle\langle
T\rangle}\right)
\end{eqnarray*}
so that, at the random fixed point
\begin{displaymath}
\overline{\langle TT\rangle_c}=c_{\rm eff}/2z^4
\end{displaymath}
where $c_{\rm eff}=c'(0)$, and
\begin{eqnarray*}
\delta c_{\rm eff}&=&-3\pi\lambda^2(1-2\Delta)^2\lim_{n\to0}(1/n)\\&&
\sum_{a\not=b}\sum_{c\not=d}\int r^2\langle\Phi_a(r)\Phi_b(r)\Phi_c(0)
\Phi_d(0)\rangle_cd^2\!r\\
&=&-3\pi(1-2\Delta)^2\int r^2
\overline{h(r)h(0)\langle\Phi(r)\Phi(0)\rangle_c}d^2\!r\\
&=&
-{3\pi(1-2\Delta)^2\over{\rm area}}\int r_{12}^2
h(r_1)h(r_2)\langle\Phi(r_1)\Phi(r_2)\rangle_cd^2\!r_1d^2\!r_2
\end{eqnarray*}
The seocnd expression follows from undoing the replacement\break
$h\to\lambda\sum_a\Phi_a$ in the gaussian integration.
The last line expresses the fact that the quenched average is
unnecessary if instead we average over the whole system: this
version of the $c$-theorem sum rule thus applies to a \em 
given realisation \em of the randomness.
Note there is no obvious positivity:  we expect that \
$h\Phi>0$, but the above involves
$h(\Phi-\langle\Phi\rangle)$.

However, there are in addition $n-1$ other independent components of the 
deformed stress tensor: 
\begin{eqnarray*}
{\cal T}&=&\sum_a T_a\\
\Tt_a&=& T_a-(1/n){\cal T}
\end{eqnarray*}
where $\sum_a\Tt_a=0$. 
These combinations are chosen to transform according to
irreducible representations of ${\rm S}_n$, so that they
should deform into conformal fields at the new
fixed point, with well-defined scaling dimensions
$(2+\delta(n),\delta(n))$. It may be checked in perturbation theory that
$\delta\not=0$.

In the undeformed theory, 
\begin{displaymath}
\langle \Tt_a\Tt_b\rangle=\left(\delta_{ab}-{1\over n}\right){c\over
2z^4}
\end{displaymath}
so we choose, at the new fixed point,
\begin{displaymath}
\langle \Tt_a\Tt_b\rangle=\left(\delta_{ab}-{1\over n}\right)
{c(n)\over 2n}{1\over z^4(z\zb)^{2\delta(n)}}
\end{displaymath}
Then
\begin{eqnarray*}
\overline{\langle T\rangle\langle T\rangle}&=&
\lim_{n\to0}\langle T_1T_2\rangle\\
&=&\lim_{n\to0}\langle(\Tt_1+(1/n){\cal T})(\Tt_2+(1/n){\cal T})\rangle\\
&=&\lim_{n\to0}{c'(0)\over2z^4}\left(
-{1\over n}(z\zb)^{-2\delta(n)}+{1\over n}\right)\\
&=&{\tilde c_{\rm eff}\over 2z^4}\,\ln(z\zb)
\end{eqnarray*}
where
\begin{displaymath}
\tilde c_{\rm eff}=2c'(0)\delta'(0)
\end{displaymath}

Now $\Tt_a$ is not conserved: in fact
\begin{displaymath}
\partial_\zb\Tt_a+\partial_z\widetilde\vartheta_a=K_a
\end{displaymath}
where
\begin{eqnarray*}
\widetilde\vartheta_a&=&
-\pi\lambda(\ffrac12-\Delta)\Phi_a\sum_{c\not=a}\Phi_c\\
K_a&=&\ffrac12\pi\lambda\sum_{b\not=a}
\left(\Phi_a\partial_z\Phi_b-\Phi_b\partial_z\Phi_a\right)
\end{eqnarray*}

This is equivalent to the previous equation for a fixed $h(r)$ by
the substitution $\lambda\sum_b\Phi_b\to h(r)$.
From the above one can derive a sum rule for $\delta\tilde c_{\rm eff}$ in
terms of suitably averaged correlators of $\Phi$ (but once again there
is no positivity).

In a general renormalisation scheme we find that
\begin{displaymath}
\widetilde\vartheta_a=-\frac12\pi
\big(\beta(\lambda)+\delta(n)\big)\sum_{c\not=a}\big(\Phi_a\Phi_c\big)_R
-(1/n)\vartheta
\end{displaymath}
so that $\widetilde\vartheta_a=O(n)$ at the random fixed point.
One should, however, be cautious in setting it to zero at $n=0$ in
correlation functions, because of factors of $1/n$.

\subsection*{The torus partition function.}

For a general conformal field theory, the
torus partition function encodes its operator content.
In the replicated theory for general $n$ we expect therefore
\begin{eqnarray*}
\overline{Z^n}&=&(q\qb)^{-c(n)/24}
\left(1+q^2+(n-1)q^{2+\delta(n)}
\qb^{\delta(n)}+\qb^2+\right.\\
&&\qquad\qquad\qquad\qquad\left.+
(n-1)q^{\delta(n)}\qb^{2+\delta(n)}+
q^2\qb^2+\cdots\right)
\end{eqnarray*}
Now this should equal 1 when $n=0$. It is clear to see how the $O(q^2)$
and $O(\qb^2)$ terms cancel, but all the descendants of these must do
so in addition! It is easy to see that this requires postulating the
existence of new Virasoro primaries at each level, whose scaling
dimensions coincide with those of the descendents of more relevant
ooperators at $n=0$. 
This suggests that there is 
massive degeneracy of Virasoro primaries as $n\to0$, suggesting that
there is an underlying extended algebra in all such theories, possibly
supersymmetry, even when it is not apparent.

It is interesting to compute the
quenched free energy $\overline{\ln Z}=$
\begin{displaymath}
(\partial/\partial n)|_{n=0}\overline{Z^n}
=-(c_{\rm eff}/24)\ln(q\qb)-\delta'(0)(q^2+\qb^2)\ln(q\qb)+\cdots
\end{displaymath}
where $\delta'(0)=\tilde c_{\rm eff}/2c_{\rm eff}$.
We see the appearance of logarithms, and also the second effective
central charge $\tilde c_{\rm eff}$.
There are still many unresolved questions, including 
how modular invariance works, and how to characterise
boundary states.

\subsection*{Operator product expansions.}
Let us begin by describing the so-called
``$c\to0$ catastrophe". For any primary operator $\phi$ in any CFT,
its OPE with itself takes the form
\begin{displaymath}
\phi(z,\zb)\cdot\phi(0,0)=
{a_\phi\over z^{2\Delta}\zb^{2\bar\Delta}}
\left(1+{2\Delta\over c}z^2T+
\cdots+{4\Delta\bar\Delta\over c^2}z^2\zb^2(T\overline T)+\cdots\right)
\end{displaymath}
so, in the 4-point function,
\begin{displaymath}
\langle\phi\phi\phi\phi\rangle\propto
a_\phi^2(1+(2\Delta/c)^2(c/2)\eta^2+\cdots
+O(1/c^4)c^2(\eta\bar\eta)^2+\cdots)
\end{displaymath}
where $\eta$ is the cross-ratio.
There is an obvious problem as $c\to0$. There are three possible resolutions:
\begin{enumerate}
\item other operators in $\cdots$ cancel the divergence;
\item $a_\phi\to0$ as $c\to0$;
\item $(\Delta,\bar\Delta)\to(0,0)$ as $c\to0$.
\end{enumerate}

Let us see what happens in the replica approach.
Set $\Phi=\sum_a\Phi_a$, 
$\Phit_a=\Phi_a-(1/n)\Phi$. These are chosen to transform according
to irreducible representations of the permutation group of the
replicas. In the pure theory,
the OPEs are schematically
\begin{eqnarray*}
\Phit_a\cdot\Phit_a&=&(1-1/n)(z\zb)^{-4\Delta}
\left(1+{2\Delta\over cn}z^2{\cal T}+{2\Delta\over
c}z^2\Tt_a+\cdots\right)\\
\Phi\cdot\Phi&=&n(z\zb)^{-4\Delta}
\left(1+{2\Delta\over cn}z^2{\cal T}
+{2\Delta^2\over(cn)^2}(z\zb)^2{\cal T}\overline{\cal T}+\right.\\
&&\qquad\qquad\qquad\qquad\left.
+{2\Delta^2\over c^2}(z\zb)^2\sum_a\Tt_a\overline{\Tt}_a+\cdots\right)
\end{eqnarray*}
which deform into
\begin{eqnarray*}
\Phit_a\cdot\Phit_a&=&(1-1/n)(z\zb)^{-4\Delta_\Phit}
\left(1+{2\Delta_\Phit\over c(n)}z^2{\cal T}+\right.\\
&&\qquad\qquad\qquad\qquad\left.+{\rm const\ }z^2(z\zb)^\delta(n)
\Tt_a+\cdots\right)\\
\Phi\cdot\Phi&=&n(z\zb)^{-4\Delta_\Phi}
\left(1+{2\Delta_\Phi\over c(n)}z^2{\cal T}
+{2\Delta_\Phi^2\over c(n)^2}(z\zb)^2{\cal T}\overline{\cal T}+\right.\\\
&&\qquad\qquad\qquad\qquad\left.
+ {\rm const}(z\zb)^{2+\delta_2(n)}{\cal M}+\cdots\right)
\end{eqnarray*}
where $\cal M$ is a new primary operator with dimensions $(2+\delta_2(n),
2+\delta_2(n))$.
$\Phit$ and $\Phi$ resolve the ``$c\to0$ catastrophe''
according to schemes 1 and 2 respectively.
Their 4-point functions have the form
\begin{eqnarray*}
\langle\Phit_a\Phit_a\Phit_a\Phit_a\rangle&\sim&
1+\delta'(0)\eta^2\ln(\eta\bar\eta)+\cdots\\
\langle\Phi\Phi\Phi\Phi\rangle&\sim&
n\left(\eta^2+\cdots+\delta_2'(0)
(\eta\bar\eta)^2\ln(\eta\bar\eta)+\cdots\right)
\end{eqnarray*}
Note that the connected correlators of $\Phi$ all vanish proportional
to $c$ as $n\to0$. 

$\Phi_a\equiv\Phit_a+(1/n)\Phi$ 
and $\Phi$ are an example of a \em logarithmic pair\em:
at $c=0$
\begin{eqnarray*}
\langle\Phi_a(z,\zb)\Phi_a(0,0)\rangle&\sim&(z\zb)^{-4\Delta}\ln(z\zb)\\
\langle\Phi_a(z,\zb)\Phi(0,0)\rangle&\sim&(z\zb)^{-4\Delta}\\
\langle\Phi(z,\zb)\Phi(0,0)\rangle&=&0
\end{eqnarray*}

It turns out that {\bf Kac operators} are always
examples of the second solution
to the $c=0$ catastrophe:
\begin{itemize}
\item Def.: a Kac operator $\phi$ has scaling dimensions at some fixed
position in the Kac table for a range of $c$ including $0$.
\end{itemize}

Now only other Kac operators can appear in the OPE $\phi\cdot\phi$:
this excludes a companion of $\cal T$, which would have dimension
$(2+\delta,\delta)$, which does not appear in the Kac table.
Hence we must have resolution 2: $a_\phi\to0$ as $c\to0$.
(But note that
${\cal M}$ with dimensions $(2+\delta_2,2+\delta_2)$ does exist,
giving rise to $(\eta\bar\eta)^2\ln(\eta\bar\eta)$ terms in the
4-point function. Explicit calculations confirm this.)
If we choose $a_\phi\propto c^p$, one can show that
the $2N$-point connected correlator goes like $c^{N(p-1)+1}$,
so it is natural to take $p=1$. 
This is exactly what happens in
physical examples of percolation (($Q\to1$)-Potts model) or self-avoiding
walks (O$(n\to0)$ model), where Kac operators enter into
physical quantities only through
\em derivatives \em wrt $c$ of correlators.
This suggests that Kac operators in such $c=0$ theories are always
the partner of a (non-Kac) logarithmic operator. In the above examples
these other operators may be identified away from $c=0$.

\subsection*{Summary}
\begin{itemize}
\item
The stress tensor in a general quenched random system, with a given
distribution of impurities, satisfies
\begin{displaymath}
\partial_\zb T+\partial_z\Theta=K
\end{displaymath}
with explicit expressions for $\Theta$ and $K$.
\item
at a random fixed point, 
\begin{eqnarray*}
\overline{\langle TT\rangle_c}&=&c_{\rm eff}/2z^4\\
\overline{\langle TT\rangle}&=&(\tilde c_{\rm eff}/2z^4)\ln(z\zb)
\end{eqnarray*}
\item
there are
sum rules for the change in $c_{\rm eff}$ and $\tilde c_{\rm eff}$
along a RG trajectory betwen 2 fixed points, in terms of physically
measurable correlators.
\item
there must be a
massive degeneracy of operators at $c=0$. This suggests an
extended symmetry,
but the candidates $\Tt$ for its generators are
not holomorphic fields!
\item
some operators solve the ``$c\to0$ catastrophe'' by having connected
correlators which are all $O(c)$ -- this is true of all Kac operators --
but the \em physics \em is in the $O(c)$ term and is therefore invisible in
the theory at $c=0$. This suggests that approaches to taking the
quenched average which work exactly at $c=0$, such as supersymmetry,
cannot expose all the physics.
\end{itemize}

\end{document}